\documentclass[11pt,letterpaper]{article}
\usepackage{jheppub}
\usepackage[utf8]{inputenc}
\usepackage{amsmath}
\usepackage{tensor}
\usepackage{mathtools}
\usepackage{amsfonts}
\usepackage{amssymb}
\usepackage{tikz}
\usetikzlibrary{calc,positioning}
\usepackage{float}
\usepackage{hyperref}
\allowdisplaybreaks

\newcommand{\drm}{\mathrm{d}}

\title{On the continuum limit of Benincasa-Dowker-Glaser causal set action}
\author{Ludovico Machet,}
\author{Jinzhao Wang}
\affiliation{\small \it Institute for Theoretical Physics, ETH 8093 Z\"urich, Switzerland}
\emailAdd{ludovicomachet@gmail.com}
\emailAdd{jinzwang@phys.ethz.ch}

\abstract{
We study the continuum limit of the Benincasa-Dowker-Glaser causal set action on a causally convex compact region. In particular, we compute the action of a causal set randomly sprinkled on a small causal diamond in the presence of arbitrary curvature in various spacetime dimensions. In the continuum limit, we show that the action admits a finite limit. More importantly, the limit is composed by an Einstein-Hilbert bulk term as predicted by the Benincasa-Dowker-Glaser action, and a boundary term exactly proportional to the codimension-two joint volume. Our calculation provides strong evidence in support of the conjecture that the Benincasa-Dowker-Glaser action naturally includes codimension-two boundary terms when evaluated on causally convex regions. }

\begin{document}
\maketitle

\section{Introduction}
\label{intro}

The Causal Set Theory (CST) approach to quantum gravity claims the structure of spacetime is that of a locally finite and partially ordered set of events, whose order corresponds to the macroscopic causal relations (see \cite{Surya} for an up-to-date review). In this setup, the classical continuum spacetime is seen as a coarse grained emergent property of an underlying statistical ensemble of causal sets (causets). This point of view leads naturally to the pursue of a ``sum over histories" quantisation of the theory, with each causet realisation considered as a valid ``history". One is therefore motivated to write down a quantum partition function of the form 

\begin{equation}
    Z_\Omega=\sum_{\mathcal{C}\in\Omega}e^{\frac{1}{\hbar}S(\mathcal{C})},
\end{equation}

with $\Omega$ the sample space of causets and $S(\mathcal{C})$ an action for causets. It is therefore crucial to have an discrete causet action that approaches the Einstein-Hilbert action in the continuum limit. Successful steps towards the definition of such an action have been taken studying a family of discrete d'Alembertian operators for scalar fields on causets, first introduced in dimension $2$ by Sorkin \cite{sorkin2007does} then generalised by Dowker and Glaser \cite{DG}. In a $4$-dimensional spacetime approximated by a causet $\mathcal{C}$, for $\phi:\mathcal{C}\rightarrow \mathbb{R}$, one has the operator 

\begin{equation}
    B\phi(x)=\frac{4}{\sqrt{6}}\Big[-\phi(x)+\Big(\sum_{y\in L_0}-9\sum_{y\in L_1}+16\sum_{y\in L_2}-8\sum_{y\in L_3}\Big)\phi(y)\Big],
\end{equation}

where $L_m=L_m(x)$ denotes the set of past $m-$nearest neighbours of $x$. It was shown in \cite{BBD} that under certain assumptions this operator effectively localises and in the continuum limit one has 

\begin{equation}
    \lim_{\rho\to\infty}\frac{1}{\sqrt{\rho}}\langle B\phi(x)\rangle=\left(\square-\frac{1}{2}R(x)\right)\phi(x).
\end{equation}

By setting $\phi(x)=-2$ one gets a discrete definition of the Ricci scalar curvature. Therefore, in dimension $4$, the action for a finite causet $\mathcal{C}$ of cardinality $N$ becomes 

\begin{equation}
    S^{(4)}(\mathcal{C})=\sum_{x\in \mathcal{C}}R(x)=\frac{4}{\sqrt{6}}\Big[N-N_0+9N_1-16N_2+8N_3\Big],
\end{equation}

where $N_m$ the number of $m-$inclusive intervals in $\mathcal{C}$. We call a $m$-inclusive interval a subset $I[x,y]:=\{w\in\mathcal{C}\,|\,x\prec w\prec y\}$ and $|I[x,y]|=m$. The action is manifestly nonlocal but one expects that it should localize in the continuum limit, reproducing the classical local physics. Importantly, in the continuum limit, we know that the BDG action reduces to the classical Einstein-Hilbert action up to boundary terms~\cite{BBD}. The boundary contributions were studied for causets sprinkled on  causal diamonds in flat spacetime in \cite{Buck}. There it was shown that the action contains a boundary term coming from the null-null codimension $2$ joint of the causal diamond. Mathematically, 

\begin{equation}
    \lim_{\rho\rightarrow\infty}\frac{1}{\hbar}\Big\langle S_{BDG}^{(d)}\Big\rangle=\frac{1}{l_p^{d-2}}\mathrm{vol}\left(\mathcal{J}^{(d-2)}\right),
\end{equation}

where $l_p$ is the Planck length. This is in agreement with a conjecture from Dowker and Benincasa stating that if one assumes the action indeed localizes in the continuum limit, it should only receive boundary contributions proportional to the joint's volume when evaluated on a causal diamond. The detailed arguments in favour of this conjecture will appear in a paper in preparation by Dowker \cite{Dowker}. Roughly speaking, the causal sets' points sampled close to the null boundaries of the causal diamond will always have enough spacetime in their future or past to contribute as $R$ in the d'Alembertian expectation value. Together with the order reversal symmetry of the action one concludes the null boundaries give no boundary terms. Only the points close to both the past and future boundary, i.e. close to the joint, will yield a boundary term contribution. By dimensional and locality arguments, one expects this term to be proportional to the joint volume and nothing else.\\

In this work we aim to verify the conjecture on a small causal diamond $I[p,q]$ in arbitrary spacetimes with curvature, where the curvature scale is much larger than the interval size. In section \ref{sec1} we compute the average abundance of $m$-inclusive intervals for a causet sprinkled on such region, under the assumption that the diamond is small with respect to the geometric curvature scale. In section \ref{sec2} we use this result to infer the continuum limit of the causet action. We show that the limit is compatible with the perturbation of the joint volume induced by the curvature, yielding a boundary term exactly proportional to the  joint volume.  This supports the claim that the BDG action is indeed a well localized in the continuum limit. In what follows, we will use Riemann Normal Coordinates (RNC) expansions above a flat background. We will denote quantities computed in Minkowski spacetime with a $f$ subscript. \\

\section{Counting of $m$-inclusive intervals in curved spacetime}
\label{sec1}

We are hereby interested in computing the average number of $m$-inclusive intervals $\langle N^d_m(\rho,V)\rangle$ in a causal set $\mathcal{C}$ sprinkled with a Poisson distribution onto a causal diamond of proper time $\tau$. Following the derivation presented in \cite{GS} for flat spacetime, we consider the interval $I[p,q]$ with proper time $\tau$ between $p$ and $q$. From the Poisson sprinkling process of density $\rho$, the probability that two points $(x,y)$ in $I[p,q]$ are causally linked is given by 

\begin{equation}
    P_{xy}=e^{-\rho V_{xy}}.
\end{equation}

Moreover, the probability of finding $m$ elements of $\mathcal{C}$ into the volume $V_{xy}$ between $x$ and $y$ is given by

\begin{equation}
    P_{xy}(m)=\frac{(\rho V_{xy})^m}{m!}e^{-\rho V_{xy}}.
\end{equation}

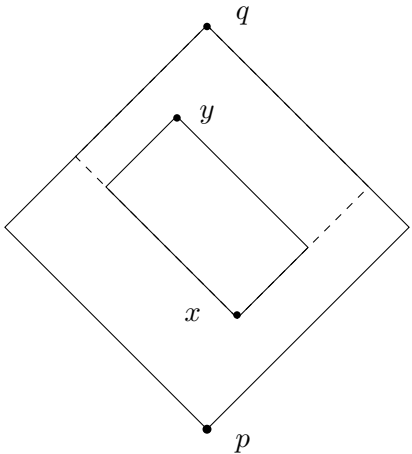
\begin{figure}[h]
\centering 
\begin{tikzpicture}[scale=1.9]
\draw[-,rotate around={45:(0,0)}] (0,0) rectangle (2,2);
\draw[dashed,rotate around={45:(0,0)}] (0.7,0.4) rectangle (2,2);
\draw[-,rotate around={45:(0,0)}] (0.7,0.4) rectangle (1.4,1.7);

\node at (0,0)[circle,fill,inner sep=1.2pt]{};
\node at (0,2.82)[circle,fill,inner sep=1pt]{};
\node at (0,0)[circle,fill,inner sep=1pt]{};
\node at (0.21,0.8)[circle,fill,inner sep=1pt]{};
\node at (-0.21,2.18)[circle,fill,inner sep=1pt]{};

\node at (0.25,-0.1) {$p$};
\node at (-0.1,.8) {$x$};
\node at (0,2.2) {$y$};
\node at (0.25,2.9) {$q$};

\end{tikzpicture}
\caption{The causal diamond $I[p,q]$ considered as integration region.}
\label{ACD}
\end{figure}

We thus consider a interval $I[p,q]$ of proper time $\tau$, with $x,y$ two elements of this region, defining an interval $I[x,y]\subset I[p,q]$, with $y$ in the future of $x$, i.e. $y\in I[x,q]$ (see figure \ref{ACD}). 

We work in a general curved spacetime and we assume the curvature scale is much larger than the proper time between $p$ and $q$ and also much larger then the discreteness scale of the causet sprinkling. This requirement is necessary in order to expand the quantities of interest in powers of the curvature tensors and scalars.\\

We shall thus evaluate the following integral

\begin{equation}
    \Big\langle N^d_m(\rho,V)\Big\rangle=\rho^2\int_{I[p,q]}\drm^dx\sqrt{-g_x}\int_{I[x,q]}\drm^dy\,\sqrt{-g_y}\,\frac{(\rho V_{xy})^m}{m!}e^{-\rho V_{xy}}.
    \label{Nm}
\end{equation}

In doing this computation, one has to be particularly careful in evaluating the corrections to the flat spacetime integral introduced by curvature. We will expand the integral at the first order in the curvature geometric quantities, generating a family of terms that will in the end sum up to give a correction to the flat spacetime action. We will then discuss the behaviour of this correction in the continuum limit, with emphasis on its scaling with respect to the number of sprinkled points and on its value with respect to the codimension two joint.\\

As already outlined, we will work in RNC and we will usually use two sets of coordinates. The first will be centred in the middle of the $I[p,q]$ diamond and will be denoted by $\left\{x^\mu\right\}$. The second will be adapted to the inner diamond $I[x,q]$ and denoted by $\left\{y^\mu\right\}$. When dealing with both coordinate systems, a ' will indicate quantities defined on $\left\{y^\mu\right\}$. Moreover, a $\Delta$ will indicate the deviations from Minkowski values, in general proportional to geometric quantities as $R$ or $R_{\mu\nu}$ components.

\subsection{Two dimensions}

We will now introduce the techniques used to tackle the action evaluation studying a sprinkling on a 2-dimensional manifold. Let us consider a spacetime $(\mathcal{M},g)$ with $g_{\mu\nu}$ a generic metric tensor, the Riemann curvature tensor can be written in the form

\begin{equation}
R_{\mu\nu\alpha\beta}=R\,g_{\mu[\alpha}g_{\beta]\nu}
\end{equation}

with $R=R(0)$ evaluated at the origin, and it has only one non-trivial component, namely $R_{0101}$. This turns out to be $R_{0101}=-R/2$. Therefore, one has \cite{Myrheim},

\begin{equation}
\sqrt{-g}=1-\frac{1}{6}R_{\mu\nu}(0)x^\mu x^\nu+\mathcal{O}(x^3)= 1+\frac{R}{12}\left(t^2-r^2\right)+\mathcal{O}(x^3).
\end{equation}

Moreover, the proper time between two points of coordinates $(t_x,r_x)$ and $(t_y,r_y)$, after some algebra analogous to the one presented in appendix for the higher dimensions case, is given by

\begin{align}
\tau_{xy}^2&=-g_{\mu\nu}(x-y)^\mu(x-y)^\nu\nonumber\\
&=\left(\left(t_x-t_y\right)^2-\left(r_x-r_y\right)^2\right)-\frac{R}{6}\left(r_xt_y-r_yt_x\right)^2+\mathcal{O}(x^5)\nonumber,\\
&=\tau_{xy,f}^2-\frac{R}{6}\left(r_xt_y-r_yt_x\right)^2+\mathcal{O}(x^5),
\end{align}

with $\tau_{xy,f}$ the Minkowski proper time. The causal diamond volume expansion worked out in general dimension in \cite{GibS} reduces to the following form after identifying $R_{00}=-R/2$,

\begin{equation}
V(\tau)=2\left(\frac{\tau}{2}\right)^2-\frac{1}{6}R\left(\frac{\tau}{2}\right)^4+\mathcal{O}(\tau^6).
\label{volume_conformal}
\end{equation}

Finally, one has to take into account the corrections to the light cones boundaries, which are in this case given by

\begin{align}
r_x^\pm(t_x)&=\left(\frac{\tau}{2}\mp t_x\right)\left(1+\frac{R_{00}}{6}\left(\frac{\tau}{2}\right)^2\right)+\mathcal{O}(\tau^4)\nonumber\\
&=\left(\frac{\tau}{2}\mp t_x\right)\left(1-\frac{R}{12}\left(\frac{\tau}{2}\right)^2\right)+\mathcal{O}(\tau^4).
\end{align}

We are now ready to tackle the evaluation of equation \eqref{Nm}. Setting $m=0$ and starting with the inner integral, we can choose a coordinate system such that $x=(-\tau'/2,0)$ and $q=(\tau'/2,0)$. Evaluating the RNC expansions on this coordinate system is equivalent with the expansions evaluated at the origin of the outer coordinate system at this order in curvature. Thus, 

\begin{align}
&\int_{I[x,q]}\drm^2y\, \sqrt{-g_y}\,V_{xy}^n\nonumber\\
&=\int_{I[x,q]}\drm^2y\, \left(1+\frac{R}{12}(t_y^2-r_y^2)\right)\left(\frac{\tau_{xy}^{2n}}{2^n}\right)\left(1-\frac{R}{48}\tau_{xy,f}^2\right),\nonumber\\
&=\int_{I[x,q],f}\drm^2y\,\frac{\tau_{xy,f}^{2n}}{2^n}+\int_{\Delta I[x,q]}\drm^2y\,\frac{\tau_{xy,f}^{2n}}{2^n}-\int_{I[x,q],f}\drm^2y\,\frac{nR}{6}\frac{\tau_{xy,f}^{2(n-1)}}{2^n}\left(\frac{\tau_f'}{2}r_y\right)^2,\nonumber\\
&\quad+\int_{I[x,q],f}\drm^2y\,\frac{\tau_{xy,f}^{2n}}{2^n}\left(\frac{R}{12}(t_y^2-r_y^2)-\frac{nR}{48}\tau_{xy,f}^2\right).
\end{align}

In the last set of equations, we denoted by $I[p,q],f$ the flat spacetime causal diamond between $p$ and $q$, in particular with Minkowski light cone boundaries, and by $\Delta I[x,q]$ the region between the Minkowski light cone boundaries and the generic curvature ones. \\

We will change to null coordinates to perform the integrations on $I[x,q],f$. We will also use that

\begin{equation}
\int_a^{a+\varepsilon}\mathrm{d}r\,f(r)=f(a)\varepsilon+\mathcal{O}(\varepsilon^2)
\end{equation}

in order to approximate the integral over the $\Delta I[x,q]$ region to the required order. Then, one has 

\begin{equation}
\int_{I[x,q]}\drm^2y\, \sqrt{-g_y}\,V_{xy}^n=\frac{2^{-1-n}}{(n+1)^2}\tau'^{2n+2}-R\frac{2^{-5 - n} (4 + n + n^2) }{ 3 (1 + n) (2 + n)^2}\tau_f'^{4 + 2 n}.
\end{equation}

Carrying on the same reasoning to the outer integral, we have

\begin{align}
&\int_{I[p,q]}\drm^2x\, \sqrt{-g_x}\,\left(\frac{2^{-1-n}}{(n+1)^2}\tau'^{2n+2}+R\frac{2^{-5 - n} (4 + n + n^2) }{ 3 (1 + n) (2 + n)^2}\tau_f'^{4 + 2 n}\right),\nonumber\\
&=\int_{I[p,q],f}\drm^2x\,\frac{2^{-1-n}}{(n+1)^2}\tau_f'^{2n+2}+\int_{\Delta I[p,q]}\drm^2x\,\frac{2^{-1-n}}{(n+1)^2}\tau_f'^{2n+2}-\int_{I[p,q],f}\drm^2x\,\frac{R}{6}\frac{\tau_f'^{2n}}{2^{n+1}(n+1)}\left(\frac{\tau_f}{2}r_x\right)^2\nonumber\\
&\quad+\int_{I[p,q],f}\drm^2x\,\frac{R}{12}\frac{2^{-1-n}}{(n+1)^2}\tau_f'^{2n+2}(t_x^2-r_x^2)-\int_{I[p,q],f}\drm^2x\,R\frac{2^{-5 - n} (4 + n + n^2) }{ 3 (1 + n) (2 + n)^2}\tau_f'^{4 + 2 n}\nonumber,\\
&=\frac{2^{-2-n}}{(n+1)^2(n+2)^2}\tau^{2n+4}-R\frac{2^{-6-n}(12 + n (9 + n (2 + n)))}{
	3 (1 + n)^2 (2 + n)^2 (3 + n)^2}\tau_f^{2n+6}.
\end{align}

Inserting this result in equation \eqref{Nm} and rewriting it as a function of $N_f$,

\begin{equation}
N_f=\rho\left(2\left(\frac{\tau}{2}\right)^2\right), 
\label{Nf_2d}
\end{equation}

one gets the expected number of links in the causal diamond, given by 

{\small
\begin{equation}
 \Big\langle N^{(2)}_0(\rho,V)\Big\rangle=\sum_{n=0}^\infty\left\{\frac{(-N_f)^{n+2}}{n!}\frac{1}{(n+1)^2(n+2)^2}-\frac{(-N_f)^{n+2}}{n!}R\frac{12 + n (9 + n (2 + n))}{
 12 (1 + n)^2 (2 + n)^2 (3 + n)^2}\left(\frac{\tau_f}{2}\right)^2\right\}.
\label{N0_2d}
\end{equation}
}%

By Lorentz invariance one can take the $m$th derivative of equation \eqref{N0_2d} in order to get the expected value of $m-$inclusive intervals. This reads

\begin{align}
\Big\langle N^{(2)}_m(\rho,V)\Big\rangle=&\sum_{n=0}^\infty\Bigg\{\frac{(-N_f)^{n}N_f^{m+2}}{n!m!}\frac{1}{(n+m+1)^2(n+m+2)^2}\nonumber\\
&-\frac{(-N_f)^{n}N_f^{m+2}}{n!m!}R\frac{12 + (n+m) (9 + (n+m) (2 + (n+m)))}{
	12 (1 + n+m)^2 (2 + n+m)^2 (3 + n+m)^2}\left(\frac{\tau_f}{2}\right)^2\Bigg\}.
\label{Nm_2d}
\end{align}

\subsection{Higher dimensions}
\label{RNC_d>2}

We will now study equation \eqref{Nm} in a $d$-dimensional spacetime $(\mathcal{M},g)$. The region of interest will be the causal diamond $I[p,q]$ of length $\tau$, equipped with a Riemann Normal Coordinate system centred in the middle of the time-like interval between the extremal points and such that $p^\mu=(-\tau/2,0,...,0)$ and $q^\mu=(\tau/2,0,...,0)$.  We can then start evaluating the corrections to the flat spacetime case induced by curvature. \\

The first one is obviously the presence of the square root of metric determinant in the covariant volume element. Working in the $(-,+,+,+)$ metric signature convention and with Riemann Normal Coordinates $\{x^\mu\}$ with origin in the centre of $I[p,q]$, one has \cite{Myrheim},

\begin{equation}
    \sqrt{-g}=1-\frac{1}{6}R_{\mu\nu}(0)x^\mu x^\nu+\mathcal{O}(x^3). 
\end{equation}

We define the proper times between the integration limits as follows:
\begin{align}
    \tau^2&=-g_{\mu\nu}(q-p)^\mu(q-p)^\nu,\nonumber\\
    \tau'^2&=-g_{\mu\nu}(q-x)^\mu(q-x)^\nu,\nonumber\\
    \tau_{xy}^2&=-g_{\mu\nu}(y-x)^\mu(y-x)^\nu.
    \label{PT}
\end{align}

In the RNC $\{x^\mu\}$ the metric tensor can be expanded as 

\begin{equation}
    g_{\mu\nu}=\eta_{\mu\nu}-\frac{1}{3}R_{\mu\alpha\nu\beta}x^\alpha x^\beta +\mathcal{O}(x^3).
    \label{metric}
\end{equation}

Thus, at first order in the curvature, the proper times in equation \eqref{PT} will be affected as, for example,

\begin{equation}
    \tau_{xy}^2=-\eta_{\mu\nu}(y-x)^\mu(y-x)^\nu+\frac{1}{3}R_{\mu\alpha\nu\beta}y^\alpha y^\beta(y-x)^\mu(y-x)^\nu.
\end{equation}

Furthermore, we shall consider the variation to the flat causal diamond volume $V_{xy}$ computed in \cite{Myrheim, KS, GibS}, which at the first order in the curvature is given by

\begin{align}
    V&=V_{f}(\tau)\left(1-\frac{d}{24(d+1)(d+2)}R\tau^2+\frac{d}{24(d+1)}R_{\mu\nu}(y-x)^\mu(y-x)^\nu+\mathcal{O}(x^3)\right),
\end{align}

with 

\begin{equation}\label{eq:flatvolume}
    V_{f}(\tau)=\frac{\Omega_{d-2}2^{1-d}}{d(d-1)}\tau^d=:\zeta_0\tau^d,
\end{equation}

where $\Omega_d$ is the volume of the unit sphere $S^d$, given by $\Omega_{d}=2\pi^\frac{d+1}{2}/\Gamma\left(\frac{d+1}{2}\right)$.\\

Finally, the presence of curvature affects the diamond boundaries. In particular, the radial direction gets a contribution already at the first order in the Ricci tensor, whereas the time direction is affected only at higher orders. As computed in \cite{Jinzhao}, one can express the new boundaries as 

\begin{equation}
    r^\pm(t,n)=\left(\frac{\tau}{2}\mp t\right)\left(1+\frac{R_{0i0j}n^in^j}{6}\left(\frac{\tau}{2}\right)^2+\mathcal{O}(\tau^4)\right).
    \label{limit_r}
\end{equation}

We are now ready to tackle the computation. The integration regions will be the outer interval $I[p,q]$ of proper time $\tau$, the inner causal region defined by the intersection of the future of $x$ with the outer diamond, $I[x,q]$ of length $\tau'$, and the innermost diamond $I[x,y]$ of length $\tau_{xy}$.\\

We will start the computation evaluating equation \eqref{Nm} for $m=0$, i.e. counting the number of links in the interval $I[p,q]$. This will allow us to keep the notation clean and easy to follow. The general form of $N_m^{(d)}$ will be again obtained from the number of links by differentiation with respect to $\rho$. We shall first of all evaluate the $y$ integral. Thus, considering the expansions described earlier, we have

\begin{align}
&\int_{I[x,q]}\drm^dy\,\sqrt{-g(y)}e^{-\rho V_{xy}}=\sum_{n=0}^\infty\frac{1}{n!}\int_{I[x,q]}\drm^dy\,\sqrt{-g(y)}\left(-\rho V_{xy}\right)^n,\nonumber\\
=&\sum_{n=0}^\infty\frac{(-\rho\zeta_0)^n}{n!}\int_{I[x,q]}\drm^dy\,\left(1-\frac{1}{6}R_{\mu\nu}y^\mu y^\nu\right)\tau_{xy}^{dn}\Bigg(1-\frac{d}{24(d+1)(d+2)}R\tau_{xy}^2\nonumber\\
&\hspace{2cm}+\frac{d}{24(d+1)}R_{\mu\nu}(y-x)^\mu(y-x)^\nu\Bigg)^n\nonumber,\\
=&\sum_{n=0}^\infty\frac{(-\rho\zeta_0)^n}{n!}\int_{I[x,q]}\drm^dy\,\tau_{xy,f}^{dn}\Bigg(1-\frac{1}{6}R_{\mu\nu}y^\mu y^\nu+\frac{dn}{6}\left(\frac{\tau_f'}{2}\right)^2r_y^2\tau_{xy,f}^{-2}R_{i0j0}n^in^j\nonumber\\
&\hspace{2cm}-\frac{dn}{24(d+1)(d+2)}R\tau_{xy,f}^{2}+\frac{dn}{24(d+1)}R_{\mu\nu}(y-x)^\mu(y-x)^\nu\Bigg),\nonumber\\
=&\sum_{n=0}^\infty\frac{(-\rho\zeta_0)^n}{n!}\Bigg\{\int_{I[x,q],f}\drm^dy\,\tau_{xy,f}^{dn}\tag{a}\label{eq:28.1}\nonumber\\
&\hspace{1.9cm}+\int_{\Delta I[x,q]}\drm^dy\,\tau_{xy,f}^{dn}\tag{b}\label{eq:28.2}\nonumber\\
&\hspace{1.9cm}-\int_{I[x,q],f}\drm^dy\,\tau_{xy,f}^{dn}\frac{1}{6}R_{\mu\nu}y^\mu y^\nu\tag{c}\label{eq:28.3}\nonumber\\
&\hspace{1.9cm}+\int_{I[x,q],f}\drm^dy\,\tau_{xy,f}^{dn-2}\frac{dn}{6}\left(\frac{\tau'}{2}\right)^2r_y^2R_{i0j0}n^in^j\tag{d}\label{eq:28.4}\nonumber\\
&\hspace{1.9cm}+\int_{I[x,q],f}\drm^dy\,\tau_{xy,f}^{dn}\,\frac{dn}{24(d+1)}\Bigg(-\frac{R\tau_{xy,f}^{2}}{(d+2)}+R_{\mu\nu}(y-x)^\mu(y-x)^\nu\Bigg)\Bigg\}\tag{e}\label{eq:28.5}\nonumber\\
\label{dy_RNC}
\end{align}

where we Taylor expanded the exponential function and $(1+\alpha x)^n$ around $x\simeq 0$. In the last steps we split the integral domains in flat spacetime $I[x,q],f$ and curvature contribution $\Delta I[x,q]$. In what follows we will quickly describe the computations techniques and give the contributions of the terms appearing in equation \eqref{dy_RNC}. More details on the expansion and integrals will be given in Appendix. \\

On the inner integration domain we define a set of orthonormal RNC $\{y^\mu\}$ centred at $O'$, in the middle of the $I[x,q]$ interval, and oriented along the $(q-x)^\mu$ vector, such that $y^\mu=(t_y,r_yn^i)$, with $t_y\in[-\frac{\tau'}{2},\frac{\tau'}{2}]$. The $n^i$ are radial space-like vectors normalised such that $n_in^i=1$ and $r_x$, $r_y$ are given by equation \eqref{limit_r} in the respective domains. See Figure \ref{RNcoordinates} for an illustration.\\

Term \eqref{eq:28.1} gives the flat spacetime contribution already computed in \cite{GS}, evaluated on the diamond $I[x,q]$. This term evaluates to

\begin{align}
\int_{I[x,q],f}\drm^dy\,\tau_{xy,f}^{dn}=\Omega_{d-2}\frac{n2^{-d}\Gamma(d-1)\Gamma\left(\frac{dn}{2}\right)}{(n+1)\Gamma\left(d+\frac{dn}{2}\right)}\tau'^{d(n+1)}.
\end{align}

The second term, \eqref{eq:28.2}, encodes the contribution coming from the perturbations to the light cones boundaries $\Delta I[x,q]$. It can be approximated Taylor expanding the integral as follows: 

\begin{equation}
\int_a^{a+\varepsilon}\mathrm{d}r\,f(r)=f(a)\varepsilon+\mathcal{O}(\varepsilon^2)
\label{Taylor}
\end{equation}

where $a=r^\pm(t,n)$ and 

\begin{equation}
\varepsilon^\pm=\left(\frac{\tau}{2}\mp t\right)\frac{R_{0i0j}n^in^j}{6}\left(\frac{\tau}{2}\right)^2.
\end{equation}

It evaluates to 

\begin{align}
\int_{\Delta I[x,q]}\drm^dy\,\tau_{xy,f}^{dn}=\Omega_{d-2}\frac{2^{-3-d}\Gamma(d)\Gamma\left(1+\frac{dn}{2}\right)}{3(d-1)\Gamma\left(1+d+\frac{dn}{2}\right)}\tau_f'^{d(n+1)+2}R_{0'0'}.
\end{align}

We denoted by $R_{0'0'}$ the time-time component of the Ricci tensor evaluated in the $\{y^\mu\}$ orthonormal frame, whose time direction is tangent at the origin to the timelike geodesic running from $x$ to $q$.\\

Moving forward in the expansion, \eqref{eq:28.3} comes from the volume form in RNC. It gives a term proportional to the Ricci scalar as well as a term proportional to $R_{0'0'}$:

\begin{align}
-\int_{I[x,q],f}\drm^dy\,\tau_{xy,f}^{dn}\frac{1}{6}R_{\mu\nu}y^\mu y^\nu=-\Omega_{d-2}&\Bigg(\frac{2^{-4-d}(4+n(2+dn))\Gamma(d-1)\Gamma\left(1+\frac{dn}{2}\right)}{3(n+1)\Gamma\left(2+d+\frac{dn}{2}\right)}R_{0'0'}\nonumber\\
&+\frac{2^{-2-d}\Gamma(d+1)\Gamma\left(1+\frac{dn}{2}\right)}{3(d-1)(2+d+dn)\Gamma\left(2+d+\frac{dn}{2}\right)}R\Bigg)\tau_f'^{d(n+1)+2}.
\end{align}

Term \eqref{eq:28.4} is issued from the proper time $\tau_{xy}$ expansion and it evaluates to

\begin{align}
\int_{I[x,q],f}\drm^dy\,\tau_{xy,f}^{dn-2}\frac{dn}{6}\left(\frac{\tau_f'}{2}\right)^2r_y^2R_{i0j0}n^in^j=\Omega_{d-2}\frac{2^{-4-d}n\Gamma(d+1)\Gamma\left(\frac{dn}{2}\right)}{3(d-1)(n+1)\Gamma\left(1+d+\frac{dn}{2}\right)}\tau_f'^{d(n+1)+2}R_{0'0'}.
\end{align}

Finally, term \eqref{eq:28.5} comes from the volume $V_{xy}$ perturbation, and contributes as 

\begin{align}
&\int_{I[x,q],f}\drm^dy\,\tau_{xy,f}^{dn}\Bigg(-\frac{dn}{24(d+1)(d+2)}R\tau_{xy,f}^{2}+\frac{dn}{24(d+1)}R_{\mu\nu}(y-x)^\mu(y-x)^\nu\Bigg)\nonumber\\
=&\,\Omega_{d-2}\Bigg(\frac{2^{-4-d}dn(2+d+dn)\Gamma(d)\Gamma\left(1+\frac{dn}{2}\right)}{3(d^2-1)\Gamma\left(2+d+\frac{dn}{2}\right)}R_{0'0'}\nonumber\\
&-\frac{2^{-4-d}(4+d(2+4n+d(-1+n(n+2))))\Gamma(d)\Gamma\left(1+\frac{dn}{2}\right)}{3(d^2-1)(d+2)(2+d+dn)\Gamma\left(2+d+\frac{dn}{2}\right)}R\Bigg)\tau_f'^{d(n+1)+2}.
\end{align}

After combining these terms, the inner integral evaluates to 

{\small
\begin{align}
&\int_{I[x,q]}\drm^dy\,\sqrt{-g(y)}e^{-\rho V_{xy}}=\sum_{n=0}^\infty\frac{(-\rho\zeta_0)^n}{n!}\Omega_{d-2}\Bigg\{\frac{n2^{-d}\Gamma(d-1)\Gamma\left(\frac{dn}{2}\right)}{(n+1)\Gamma\left(d+\frac{dn}{2}\right)}\tau_f'^{d(n+1)}\nonumber\\
&+\frac{2^{-4 - d} (2 (2 + n) + d (4 + n + n^2))
\Gamma(1 + d) \Gamma\left(1 + (d n)/2\right)}{3 ( d^2-1) \Gamma\left(
2 + d + \frac{dn}{2}\right)}\tau_f'^{d(n+1)+2}R_{0'0'}\nonumber\\
&-\frac{2^{-4 - 
d} (4 (2 + n) + 2 d (6 + n + 2 n^2) + 
d^2 (4 + n (-1 + n (2 + n))))
\Gamma(1 + d) \Gamma\left(1 + \frac{dn}{2}\right)}{
3 ( d^2-1) ( d+2) (2 + d + d n) \Gamma\left(2 + d + \frac{dn}{2}\right)}\tau_f'^{d(n+1)+2}R	\Bigg\}.
\label{dy_final_RNC}
\end{align}
}%

We shall now integrate equation \eqref{dy_final_RNC} on the interval $I[p,q]$. In doing so, we will again consider the expansions presented at the beginning of this section, plus the rotation required to align $R_{0'0'}$ with the time-direction defined by the geodesic running from $p$ to $q$, on which the external RNC are defined. We will indeed consider the set of RNC $\{x^\mu\}$ centred in the midpoint of the $p$-$q$ interval, such that $x^\mu=(t_x, r_xn^i)$, with $t_x\in[-\frac{\tau}{2},\frac{\tau}{2}]$ and the $n^i$ are radially directed vectors defined as before. Thus, we have 

{\small
\begin{align}
&\sum_{n=0}^\infty\frac{(-\rho\zeta_0)^n}{n!}\Omega_{d-2}\Bigg\{\frac{n2^{-d}\Gamma(d-1)\Gamma\left(\frac{dn}{2}\right)}{(n+1)\Gamma\left(d+\frac{dn}{2}\right)}\int_{I[p,q]}\drm^dx\,\sqrt{-g(x)}\tau'^{d(n+1)}\nonumber\\
+&\frac{2^{-4 - d} (2 (2 + n) + d (4 + n + n^2))
	\Gamma(1 + d) \Gamma\left(1 + (d n)/2\right)}{3 ( d^2-1) \Gamma\left(
	2 + d + \frac{dn}{2}\right)}\int_{I[p,q],f}\drm^dx\,\tau_f'^{d(n+1)+2}R_{0'0'}\nonumber\\
-&\frac{2^{-4 - 
		d} (4 (2 + n) + 2 d (6 + n + 2 n^2) + 
	d^2 (4 + n (-1 + n (2 + n))))
	\Gamma(1 + d) \Gamma\left(1 + \frac{dn}{2}\right)}{
	3 ( d^2-1) ( d+2) (2 + d + d n) \Gamma\left(2 + d + \frac{dn}{2}\right)}\int_{I[p,q],f}\drm^dx\,\tau_f'^{d(n+1)+2}R\Bigg\},
\label{dx_RNC}
\end{align}
}%

where again $I[p,q],f$ and $\tau_f'$ indicate the Minkowski interval between $p$ and $q$ and the proper time from $x$ to $q$ respectively. Evaluating this integral one needs to rotate the $\left\{y^\mu\right\}$ time direction into the $\left\{x^\mu\right\}$ coordinate system. In particular the time directions defined by these charts needs to be aligned and $R_{0
'0'}$ becomes a combination of $R_{0
0}$ and $R$, i.e.

\begin{equation}
    R_{0'0'}=\frac{1}{\tau'^2}\left(\frac{\tau}{2}-t_x\right)^2R_{00}+\frac{1}{\tau'^2}r_x^2n^in^jR_{ij}.
\end{equation}

The derivation of this equation and the integral \eqref{dx_RNC} will be detailed out in appendix. Equation \eqref{dx_RNC} evaluates to 

\begin{align}
\left\langle N_0^{(d)}\right\rangle=&\sum_{n=0}^\infty\frac{\left(-N_f\right)^{n+2}}{n! }\Bigg\{\frac{\Gamma(1 + d)^2 \Gamma\left(\frac{d}{2} ( n+1)\right) \Gamma\left(1 +\frac{dn}{2}\right)}{4 \Gamma\left(
\frac{d}{2} (n+3)\right) \Gamma\left(1 + d + \frac{dn}{2}\right)}\nonumber\\
+&\frac{d^4 (2 (2 + n) (3 + n) + d (12 + n (9 + n (2 + n)))) \Gamma(
d)^2 \Gamma\left(\frac{d}{2} ( n+1)\right) \Gamma\left(1 + \frac{dn}{2}\right)}{384 ( d+1) \Gamma\left(
1 + d + \frac{dn}{2}\right) \Gamma\left(2 + \frac{d}{2} (  n+3)\right)}R_{00}\tau^2\nonumber\\
-&\Bigg[\frac{d^5 n (4 (2 + n) (3 + n) + d^2 (12 + n (6 + n + 4 n^2 + n^3)))\Gamma(
	d)^2 \Gamma\left(\frac{dn}{2} \right) \Gamma\left( \frac{dn}{2}(n+1)\right)}{1536 ( d+1)(d+2) \Gamma\left(
	2 + d + \frac{dn}{2}\right) \Gamma\left(2 + \frac{d}{2} (  n+3)\right)}\nonumber \\ 
	&+\frac{d^6 n (36 + 2 n (13 + n (3 + 2 n))) \Gamma(
	d)^2 \Gamma\left(\frac{dn}{2} \right) \Gamma\left( \frac{dn}{2}(n+1)\right)}{1536 ( d+1)(d+2) \Gamma\left(
	2 + d + \frac{dn}{2}\right) \Gamma\left(2 + \frac{d}{2} (  n+3)\right)}\Bigg]R\tau^2\Bigg\}\,,
\label{N0_RNC_final}
\end{align}

with $N_f$ given by $N_f=\rho\zeta_0\tau^d$. Let us notice the first term of this equation is precisely the Minkowski spacetime number of links and can be put in the same form given in \cite{GS} using the properties of the Gamma function. The expectation number of $m$-inclusive intervals in the $I[p,q]$ diamond can now be inferred taking $m$ derivatives of equation \eqref{N0_RNC_final} with respect to $\rho$, i.e.

\begin{equation}
    \left\langle N_m^{(d)}\right\rangle=\frac{\rho^m}{m!}\frac{\partial^m}{\partial\rho^m}\left\langle N_0^{(d)}\right\rangle\,.
    \label{Nm_RNC_final}
\end{equation}

Let us stress that $\left\langle N_m^{(d)}\right\rangle$ is expressed as a function of $N_f$, i.e. of the number of causet elements sprinkled on a diamond of length $\tau$ in Minkowski spacetime at density $\rho$.

\section{Benincasa-Dowker-Glaser action for a causal set}
\label{sec2}

Starting from the average number of $m$-inclusive intervals in a Causal Diamond of proper time $\tau$ it is possible to define an action for a causal set $\mathcal{C}$. This was proposed by Benincasa, Dowker and Glaser in \cite{BD,DG,G} and takes the following form given in \cite{Buck}

\begin{equation}
    \frac{1}{\hbar} S_{BDG}^{(d)}=\alpha_d\left(\frac{l}{l_p}\right)^{d-2}\left(N+\frac{\beta_d}{\alpha_d}\sum_{m=0}^{n_d-1}C_m^{(d)}N_m\right),
\end{equation}

where $N$ is the cardinality of the causal set, $(l/l_p)$ is the ratio of a fundamental length to the Planck length and the constants are

\begin{equation}
    \alpha_d=\begin{cases}\frac{1}{\Gamma\left(1+\frac{2}{d}\right)}c_d^{2/d}\quad d \,\,\text{odd}\\\frac{2}{\Gamma\left(1+\frac{2}{d}\right)}c_d^{2/d}\quad d \,\,\text{even},\end{cases}
\end{equation}

\begin{equation}
    \beta_d=\begin{cases}\frac{d+1}{2^{d-1}\Gamma\left(1+\frac{2}{d}\right)}c_d^{2/d}\quad\quad\, d \,\,\text{odd}\\\frac{\Gamma\left(2+\frac{d}{2}\right)\Gamma\left(1+\frac{d}{2}\right)}{\Gamma\left(\frac{2}{d}\right)\Gamma\left(d\right)}c_d^{2/d}\quad d \,\,\text{even},\end{cases}
\end{equation}

and 

\begin{equation}
    n_d=\begin{cases}\frac{d}{2}+\frac{3}{2}\quad d \,\,\text{odd}\\\frac{d}{2}+2\quad d \,\,\text{even},\end{cases}
\end{equation}

with $c_d=2^{1-\frac{d}{2}}\Omega_{d-2}/(d(d-1))$. Finally, the $C_m^{(d)}$ coefficients are

\begin{equation}
    C_m^{(d)}=\begin{cases}\sum_{k=0}^{m}(-1)^k\binom{m}{k}\frac{\Gamma\left(\frac{d}{2}(k+1)+\frac{3}{2}\right)}{\Gamma\left(\frac{d}{2}+\frac{3}{2}\right)\Gamma\left(\frac{d}{2}k+1\right)}\quad d \,\,\text{odd}\\\sum_{k=0}^{m}(-1)^k\binom{m}{k}\frac{\Gamma\left(\frac{d}{2}(k+1)+2\right)}{\Gamma\left(\frac{d}{2}+2\right)\Gamma\left(\frac{d}{2}k+1\right)}\quad d \,\,\text{even}.\end{cases}
\end{equation}

In \cite{Buck} it is shown that the leading order contribution for the BDG action on a causal diamond in the continuum limit, i.e. keeping $\tau$ fixed and sending $\rho=l^{-d}\rightarrow\infty$, for $d=2,\dots,16$ are given by

\begin{equation}
    \lim_{N\rightarrow\infty}\frac{1}{\hbar}\Big\langle S_{BDG}^{(d)}\Big\rangle=\frac{1}{l_p^{d-2}}\mathrm{vol}\left(\mathcal{J}^{(d-2)}\right)=\frac{1}{l_p^{d-2}}\Omega_{d-2}\left(\frac{\tau}{2}\right)^{d-2},
    \label{Joint_vol}
\end{equation}

with $\mathcal{J}^{(d-2)}:=\partial J^+(p)\cap\partial J^-(q)$ the codimension-$2$ joint of the causal diamond, given by the intersection of the future light-cone of $p$ with the past light-cone of $q$. This calculation for the leading contribution is carried out in Minkowski spacetime. The goal of this work discussion is to evaluate the contribution of curvature to the $S_{BDG}$ action and to study if the relation to the joint term holds at the presence of curvature. In the following, we will call $\mathcal{J}_f^{(d-2)}$ the joint of the diamond in flat spacetime. The curvature perturbation to the joint volume was evaluated in \cite{Jinzhao} and it is given by

\begin{equation}
    \delta \mathcal{J}^{(d-2)}=-\frac{\Omega_{d-2}(\tau/2)^d(R-(d-4)R_{00})}{6(d-1)}.
\end{equation}

The flat spacetime behaviour of the BDG action suggests one should expect the perturbed action to be asymptotically proportional to the deformed joint volume in the continuum limit. As an example, in the case $R=0$ and $d=4$ one expects to obtain no correction to the flat spacetime value. \\

\subsection{Two dimensions}

In dimension $2$, the BDG action takes the form 

\begin{equation}
 \frac{1}{\hbar}\Big\langle S_{BDG}^{(2)}\Big\rangle=2\left(N-2\Big\langle N_0^{(2)}\Big\rangle+4\Big\langle N_1^{(2)}\Big\rangle-2\Big\langle N_2^{(2)}\Big\rangle\right).
\end{equation}

Recall equation \eqref{Nm_2d}. Inserting it into the action definition and calling $c_m$ the coefficient multiplying the Ricci scalar, $N$ the number of causet elements sprinkled into the curved spacetime interval, we can express the action in a $2$-dimensional conformally flat spacetime as 

\begin{align}
&\frac{1}{\hbar}\Big\langle S_{BDG}^{(2)}\Big\rangle=2\left(N+\sum_{m=0}^2\alpha^{(2)}_m\Big\langle N_m^{(2)}\Big\rangle\Big\rangle\right)\nonumber,\\
&=2\left(N+\sum_{m=0}^2\alpha^{(2)}_m\sum_{n=0}^\infty\Bigg\{\frac{(-N_f)^{n}N_f^{m+2}}{n!m!}\frac{1}{(n+m+1)^2(n+m+2)^2}-\frac{(-N_f)^{n}N_f^{m+2}}{n!m!}Rc_m\left(\frac{\tau}{2}\right)^2\Bigg\}\right).
\end{align}

Equation \eqref{volume_conformal} allows to express $N_f$ as a function of $N$, i.e.

\begin{equation}
N=\rho V(\tau)=\rho\left(2\left(\frac{\tau}{2}\right)^2\right)\left(1-\frac{R}{12}\left(\frac{\tau}{2}\right)^2\right)=N_f\left(1-\frac{R}{12}\left(\frac{\tau}{2}\right)^2\right).
\end{equation}

Thus, we end up with

\begin{align}
\frac{1}{\hbar}\Big\langle S_{BDG}^{(2)}\Big\rangle&=2\Bigg(N_f\left(1-\frac{R}{12}\left(\frac{\tau}{2}\right)^2\right)\nonumber\\
&+\sum_{m=0}^2\alpha^{(2)}_m\sum_{n=0}^\infty\Bigg\{\frac{(-N_f)^{n}N_f^{m+2}}{n!m!}\frac{1}{(n+m+1)^2(n+m+2)^2}
+\frac{(-N_f)^{n}N_f^{m+2}}{n!m!}Rc_m\left(\frac{\tau}{2}\right)^2\Bigg\}\Bigg).
\end{align}

In the continuum limit, this yields

\begin{equation}
\lim_{N_C\rightarrow\infty}\frac{1}{\hbar}\Big\langle S_{BDG}^{(2)}\Big\rangle=\frac{R}{2}\left(\frac{\tau^2}{2}\right)+2.
\label{continuum_conformal}
\end{equation}

This is consistent with the behaviour one expects from the BDG action, as $R/2$ is the gravitational Lagrangian, $\tau^2/2$ the diamond volume and $2$ the joint contribution.

\subsection{Higher dimensions}

We will now focus on the cases $d=3$, $d=4$ and $d=5$, for which the action reads as follows

\begin{align}
    \frac{1}{\hbar}\Big\langle S_{BDG}^{(3)}\Big\rangle&=\alpha_3\left(\frac{l}{l_p}\right)\left(N-\Big\langle N_0^{(3)}\Big\rangle+\frac{27}{8}\Big\langle N_1^{(3)}\Big\rangle-\frac{9}{4}\Big\langle N_2^{(3)}\Big\rangle\right),\\
    \frac{1}{\hbar}\Big\langle S_{BDG}^{(4)}\Big\rangle&=\alpha_4\left(\frac{l}{l_p}\right)^2\left(N-\Big\langle N_0^{(4)}\Big\rangle+9\Big\langle N_1^{(4)}\Big\rangle-16\Big\langle N_2^{(4)}\Big\rangle+8\Big\langle N_3^{(4)}\Big\rangle\right),\\
    \frac{1}{\hbar}\Big\langle S_{BDG}^{(5)}\Big\rangle&=\alpha_5\left(\frac{l}{l_p}\right)^3\left(N-\frac{3}{8}\Big\langle N_0^{(5)}\Big\rangle+\frac{645}{128}\Big\langle N_1^{(5)}\Big\rangle-\frac{675}{64}\Big\langle N_2^{(5)}\Big\rangle+\frac{375}{64}\Big\langle N_3^{(5)}\Big\rangle\right).
\end{align}



Using equation \eqref{Nm_RNC_final} , one can write the BDG action in curved spacetime as 

\begin{align}
    \frac{1}{\hbar} \Big\langle S_{BDG}^{(d)}\Big\rangle&=\alpha_d\left(\frac{l}{l_p}\right)^{d-2}\left(N+\frac{\beta_d}{\alpha_d}\sum_{m=0}^{n_d-1}C_m^{(d)}\left(\Big\langle N_{m,f}\Big\rangle+\Delta\Big\langle N_{m}\Big\rangle\right)\right).
    \label{DeltaS}
\end{align}

One can now relate the number of points sprinkled in a flat causal diamond of volume $V_f$ to the one of points sprinkled into a causal diamond embedded in a generic spacetime, of volume $V$, i.e.

\begin{align}
     N&=\rho V=\rho V_f\left(1-\frac{R d}{24(d+1)(d+2)}\tau^2+\frac{R_{00}d}{24(d+1)}\tau^2+\mathcal{O}(\tau^4)\right).
    \label{cardinality}
\end{align}

One notices that this introduces two new correction terms proportional to $R\tau^2$ and $R_{00}\tau^2$ respectively, which one must sum to the $\Delta\Big\langle N_{m}\Big\rangle$ given by \eqref{Nm_RNC_final}. Combining eqaution \eqref{DeltaS} and \eqref{cardinality}, we are left with 

\begin{align}
    \frac{1}{\hbar} \Big\langle S_{BDG}^{(d)}\Big\rangle
    =&\alpha_d\left(\frac{l}{l_p}\right)^{d-2}\Bigg(N_f\left(1-\frac{R d}{24(d+1)(d+2)}\tau^2+\frac{R_{00}d}{24(d+1)}\tau^2\right)\nonumber\\
    &+\frac{\beta_d}{\alpha_d}\sum_{m=0}^{n_d-1}C_m^{(d)}\Bigg(\Big\langle N_{m,f}\Big\rangle+\Delta\Big\langle N_{m}\Big\rangle\Bigg)\Bigg)\nonumber,\\
    =&\frac{1}{\hbar}\Big\langle S_{BDG,f}^{(d)}\Big\rangle+\alpha_d\left(\frac{l}{l_p}\right)^{d-2}\tau^2\Bigg(R\Big\langle\Delta S_{BDG,R}^{(d)}\Big\rangle+R_{00}\Big\langle\Delta S_{BDG,R_{00}}^{(d)}\Big\rangle\Bigg),
    \label{final_S_expansion}
\end{align}

where we recognised the form of the flat spacetime action limiting to the volume of the diamond joint for $N_f\rightarrow\infty$ and collected all the curvature corrections into the $\Delta$ terms. The behaviour of $\Big\langle\Delta S^{(d)}_{BDG,R}\Big\rangle$, $\Big\langle\Delta S^{(d)}_{BDG,R_{00}}\Big\rangle$ can be evaluated numerically with \texttt{Mathematica}. The computation shows that both correction terms scale as $N^{\frac{d-2}{d}}$ and one can define the coefficients $c^{(d)}_R$ and $c^{(d)}_{R_{00}}$ as

\begin{align}
c^{(d)}_{R}:=\lim_{N\rightarrow\infty}\Big\langle\Delta S_{BDG,R}^{(d)}\Big\rangle/N^{\frac{d-2}{d}},\quad c^{(d)}_{R_{00}}:=\lim_{N\rightarrow\infty}\Big\langle\Delta S_{BDG,R_{00}}^{(d)}\Big\rangle/N^{\frac{d-2}{d}}.
    \label{Action_scaling}
\end{align}

One gets the following asymptotic behaviours for the $R$ correction

\begin{align}
    &c^{(3)}_{R}=\frac{9\sqrt{3}\pi}{1760\,\Gamma\left(-\frac{11}{3}\right)}, & &c^{(4)}_{R}=\frac{\sqrt{\pi}}{48}, & &c^{(5)}_{R}=\frac{\pi^{12/5}\,\Gamma\left(\frac{7}{5}\right)}{7680(5)^{1/5}}.
    \label{limit_R}
\end{align}

On the other hand, the correction proportional to $R_{00}$ limits to

\begin{align}
    &c^{(3)}_{R_{00}}=\frac{-9\sqrt{3}\pi}{1760\,\Gamma\left(-\frac{11}{3}\right)}, & &c^{(4)}_{R_{00}}=0, & &c^{(5)}_{R_{00}}=\frac{-625\sqrt{50-10\sqrt{5}}\pi}{753984\Gamma\left(-\frac{22}{5}\right)}.
    \label{limit_R00}
\end{align}

Let us recall, at the zero$^{\text{th}}$ order, 

\begin{equation}
    N\simeq\rho V_f(\tau)=
    l^{-d}\zeta_0\tau^d,
    \label{tau}
\end{equation}

with $\zeta_0=\frac{\Omega_{d-2}2^{1-d}}{d(d-1)}$ from (\ref{eq:flatvolume}). Using equation \eqref{tau} we have

\begin{align}
    \alpha_dl^{d-2}\tau^2\Bigg(R\Big\langle\Delta S_{BDG,R}^{(d)}\Big\rangle+R_{00}\Big\langle\Delta S_{BDG,R_{00}}^{(d)}\Big\rangle\Bigg)=&\alpha_dl^{d-2}\tau^2(c_{R}R+c_{R_{00}}R_{00})\left(l^{-d}\zeta_0\tau^d\right)^{\frac{d-2}{d}},\nonumber\\
    =&\alpha_d\zeta_0^{\frac{d-2}{d}}\tau^d(c_{R}R+c_{R_{00}}R_{00}).
\end{align}

Combining this result with equations \eqref{Action_scaling}, \eqref{limit_R} and \eqref{limit_R00}, one gets

\begin{align}
    \lim_{N\rightarrow\infty}\frac{1}{\hbar} \Big\langle S_{BDG}^{(3)}\Big\rangle=&\frac{1}{l_p}\mathrm{vol}\left(\mathcal{J}_f^{(1)}\right)+\frac{\tau^3}{l_p}\left(\frac{\pi}{48}R-\frac{\pi}{48}R_{00}\right),\nonumber\\
   \lim_{N\rightarrow\infty}\frac{1}{\hbar} \Big\langle S_{BDG}^{(4)}\Big\rangle=&\frac{1}{l_p^2}\mathrm{vol}\left(\mathcal{J}_f^{(2)}\right)+\frac{\tau^4}{l_p^2}\left(\frac{\pi}{144}R\right),\nonumber\\
   \lim_{N\rightarrow\infty}\frac{1}{\hbar} \Big\langle S_{BDG}^{(5)}\Big\rangle=&\frac{1}{l_p^3}\mathrm{vol}\left(\mathcal{J}_f^{(3)}\right)+\frac{\tau^5}{l_p^3}\left(\frac{\pi^2}{1920}R+\frac{\pi^2}{384}R_{00}\right).
\end{align}

The terms in $\mathrm{vol}(\mathcal{J}^{(d)}_f)$ are given by the flat spacetime components of the causal set action, $\frac{1}{\hbar}\Big\langle S_{BDG,f}^{(d)}\Big\rangle$ in equation \eqref{final_S_expansion}, as calculated in \cite{Buck}. One can observe that both correction terms are proportional to $\tau^d$, which allows to extract a first term proportional to the causal diamond volume, which multiplies the Ricci scalar. It is thus a good candidate for the bulk term of the action. The second term, also proportional to $\tau^d$, is of the good dimension to be a term originating from the volume of the causal diamond joint.  \\

Let us now confirm our speculation. We set $l_p=(8\pi G/\hbar)^\frac{1}{d-2}$. In the continuum, one should expect the gravitational action yields 

\begin{align}
\frac{1}{\hbar}  S^{(d)}_{\text{continuum}}=&\frac{1}{2l_p^{d-2}}RV+\frac{1}{l_p^{d-2}}\mathrm{vol}\left(\mathcal{J}_f^{(d-2)}\right)+\frac{1}{l_p^{d-2}}\delta \mathcal{J}^{(d-2)},\nonumber\\
    =&\frac{1}{l_p^{d-2}}\mathrm{vol}\left(\mathcal{J}_f^{(d-2)}\right)+\frac{1}{2l_p^{d-2}}R\zeta_0\tau^d+\frac{1}{l_p^{d-2}}\left(-\frac{\Omega_{d-2}(\tau/2)^d(R-(d-4)R_{00})}{6(d-1)}\right),\nonumber\\
    =&\frac{1}{l_p^{d-2}}\mathrm{vol}\left(\mathcal{J}_f^{(d-2)}\right)+\frac{\tau^d}{l_p^{d-2}}\left(-\frac{\Omega_{d-2}(d-6)}{2^{d+1}3d(d-1)}R+\frac{\Omega_{d-2}(d-4)}{2^{d+1}3(d-1)}R_{00}\right).
\end{align}

In particular, 

\begin{align}
    \frac{1}{\hbar}  S^{(3)}_{\text{continuum}}=&\frac{1}{l_p}\mathrm{vol}\left(\mathcal{J}_f^{(1)}\right)+\frac{\tau^3}{l_p}\left(\frac{\pi}{48}R-\frac{\pi}{48}R_{00}\right),\nonumber\\
    \frac{1}{\hbar}  S^{(4)}_{\text{continuum}}=&\frac{1}{l_p^2}\mathrm{vol}\left(\mathcal{J}_f^{(2)}\right)+\frac{\tau^4}{l_p^2}\left(\frac{\pi}{144}R\right),\nonumber\\
    \frac{1}{\hbar}  S^{(5)}_{\text{continuum}}=&\frac{1}{l_p^3}\mathrm{vol}\left(\mathcal{J}_f^{(3)}\right)+\frac{\tau^5}{l_p^3}\left(\frac{\pi^2}{1920}R+\frac{\pi^2}{384}R_{00}\right).
\end{align}

We therefore observe the expected value and the computed one agree for dimension $3,\,4$ and $5$. Our result shows that the BDG action evaluated on a small causal diamond gives, in the continuum limit, the Einstein-Hilbert term plus a boundary exactly proportional to the joint area. Thus,

\begin{equation}
\lim_{N\rightarrow\infty}\frac{1}{\hbar} \Big\langle S_{BDG}^{(d)}\Big\rangle=\frac{1}{2l_p^{d-2}}RV_f+\frac{1}{l_p^{d-2}}\mathrm{vol}\left(\mathcal{J}_f^{(d-2)}\right)+\frac{1}{l_p^{d-2}}\delta \mathcal{J}^{(d-2)}.
\end{equation}

\section{Conclusion}

In this work we compute the expected value of $m$-inclusive intervals in a causal set sprinkled on a causal diamond $I[p,q]$ with curvature. Spacetime curvature is treated at the first order with a non-null Ricci tensor which plays the role of parameter in this study. This supplies the first order correction to the flat spacetime result computed in \cite{GS}. It results a bulk term proportional to the Ricci scalar curvature and a term proportional to the $R_{00}$ component. Both these term scale correctly in the continuum limit, and the action limits is finite. The contributions from curvature sum up to give a Einstein-Hilbert bulk term and a boundary term which exactly matches the correction induced by curvature on the diamond joint volume. This result suggests the nonlocal BDG action is indeed localized in the continuum limit and it is well behaved on a generic curved spacetime. Our approach has the advantages of being valid on general Riemann Normal neighbourhoods. It fails however to give an intuitive understanding of the origins of the different terms composing the continuum limit. One cannot clearly distinguish from which part of the expansion the bulk term as well as the joint terms come from. For conformally flat spacetime, a similar calculation can be carried out in conformal coordinates where one can pin down the exactly where contributions come from \cite{Dowker}. Our result is in favour of the conjecture that the BDG action of a causally convex compact region naturally includes a codimension-two joint term in the continuum limit. One can try different calculations to support the conjecture further. Future work can be directed towards the extension of the same computation to the Weyl tensor squared order, to study the correction to the next non-null curvature order in vacuum. This will allow us to get rid of the bulk terms and have a better grasp on the origin of the boundary terms. We however think this task to be engaging, given the number of corrections one should track in the computation. One can also look into spacetime regions with more general shapes and combinations of boundaries, such as the domain of dependence of any compact spatial slice. One simple example that can be readily checked is the causal diamond truncated by a null plane intersecting it. We leave this calculation for future works. For regions that are not causally convex, things become much more difficult. The behaviour of the BDG action in the presence of timelike boundaries is, for example, still an open question from an analytical point of view.

\acknowledgments

We are grateful to Fay Dowker for valuable feedbacks and discussions on this work.
Our interactions led to an independent derivation of the main results through different methods. We concurred to publish both treatments, as we think they are both valuable for future work. We also thank William Cunningham and Sumati Surya for discussions. This work is supported by the Swiss National Science Foundation via the National Center for Competence in Research ``QSIT", and by the Air Force Office of Scientific Research (AFOSR) via grant FA9550-16-1-0245.

\appendix
\section{Computation details}

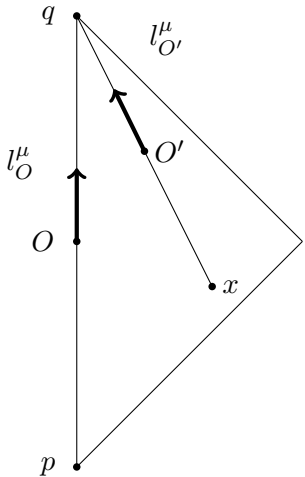
\begin{figure}[h]
\centering 
\begin{tikzpicture}[scale=1.5]

\coordinate (1) at (0,-2);
\coordinate (2) at (0,2);
\coordinate (3) at (1.2,-0.4);
\coordinate (4) at (1,0.5);
\coordinate (5) at ($(3)!.5!(2)$);

\draw (1) -- (2,0) ;
\draw (2) -- (2,0) ;

\draw[-] (1) -- (2) ;
\draw[-] (3) -- (2) ;

\draw[->, ultra thick] (0,0) -- (0,0.65);
\draw[->, ultra thick] (5) -- (0.33,1.35);

\node at (1)[circle,fill,inner sep=1pt]{};
\node at (2)[circle,fill,inner sep=1pt]{};
\node at (3)[circle,fill,inner sep=1pt]{};

\node at (5)[circle,fill,inner sep=1pt]{};
\node at (0,0)[circle,fill,inner sep=1pt]{};

\node at (-0.1,-2) [left] {$p$};
\node at (1.2,-0.4) [right] {$x$};
\node at (-0.1,2) [left] {$q$};
\node at (5) [right] {$O'$};
\node at (-0.3,0) {$O$};

\node at (-0.5,0.7) {$l^\mu_O$};
\node at (0.8,1.8) {$l^\mu_{O'}$};

\end{tikzpicture}
\caption{The causal diamond $I[p,q]$ and the time directions of $\{y^\mu\}$ and $\{x^\mu\}$ coordinates systems.}
\label{RNcoordinates}
\end{figure}

We will hereby detail out the computations performed to evaluate the integral \eqref{Nm}. As already stated, the integral is divided into two integration domains, namely $I[x,q]$ and $I[p,q]$, to which we refer as inner and outer domain respectively. On the inner region we define local Riemann Normal Coordinates $\{y^\mu\}$, with origin $O'$ and time direction given by the unit vector $l^\mu_{O'}$ tangent to the geodesic from $x$ to $q$ in $O'$, and such that $q^\mu=(\tau'/2,0,...,0)$, $x^\mu=(-\tau'/2,0,...,0)$, $y^\mu=(t_y,r_yn^i)$. The $n^i$ vectors are radially directed and normalised $n^in_i=1$. We apply the same reasoning to the outer region, defining a second set of RNC $\{x^\mu\}$ with origin $O$ and directed along $l^\mu_O$, tangent to the geodesic from $p$ to $q$ in O. We can therefore write $q^\mu=(\tau/2,0,...,0)$, $p^\mu=(-\tau/2,0,...,0)$, $x^\mu=(t_x,r_xn^i)$. We slightly abuse the notation using the same symbol $n^i$ to indicate radially directed vectors in both coordinates systems, however their occurrence in the computation will remove the ambiguity. This setup is sketched in figure \ref{RNcoordinates}.

In this frame, the curvature tensors entering the integration in the inner domain will be evaluated at $O'$ and will be contracted with the time direction given by $l^\mu_{O'}$. The tensors entering the outer domain will be evaluated at $O$ and contracted with $l^\mu_O$. The final result will contain tensors evaluated at $O$, the corrections due to their parallel transport from $O'$ to $O$ entering at higher orders in the curvature. However, a prescription is needed to express the tensors contracted with $l^\mu_{O'}$ as a function of those in the outer domain. \\

In order to do so we assume the domain of definition of RNC $\{y^\mu\}$ and $\{x^\mu\}$ to be large enough so that both contains the points $O$ and $O'$. Then, the coordinates transformation between these two systems is given by \cite{Brewin}

\begin{equation}
    y^\mu=\Lambda^\mu_\nu\left(\Delta x^\nu-\frac{1}{3}R^\nu_{\alpha\beta\rho}\Delta x^\alpha\Delta x^\beta x^\rho_{O'} \right), \label{RNC}
\end{equation}

where $\Delta x^\mu=x^\mu-x^\mu_{O'}$ and $\Lambda^\mu_\nu$ is a matrix aligning the axes of the two coordinates systems. We can therefore write

\begin{equation}
    \widetilde{l}^\mu_{O'}=\Lambda^\mu_\nu l^\nu_O,
\end{equation}

where $\widetilde{l}^\mu_{O'}$ is the vector $l^\mu_{O'}$ expressed in the $\{x^\mu\}$ coordinates, i.e. $\widetilde{l}^\mu_{O'}=\frac{1}{\tau'}(\tau/2-t_x,-r_xn^i)$. Setting $l^\mu_O=(1,0,..,0)$, one gets 

\begin{align}
    \Lambda^0_0&=\frac{1}{\tau'}\left(\frac{\tau}{2}-t_x\right), &\Lambda^i_0=-\frac{r_xn^i}{\tau'}.
\end{align}

Imposing now 

\begin{equation}
    R_{\mu'\nu'}=\Lambda^\mu_{\mu'}\Lambda^\nu_{\nu'}R_{\mu\nu}, 
\end{equation}

one gets 

\begin{equation}
    R_{0'0'}=\frac{1}{\tau'^2}\left(\frac{\tau}{2}-t_x\right)^2R_{00}-\frac{2}{\tau'^2}\left(\frac{\tau}{2}-t_x\right)r_xn^iR_{0i}+\frac{1}{\tau'^2}r_x^2n^in^jR_{ij}.
    \label{rotation}
\end{equation}

We are now ready to start the evaluation of the terms appearing in the expansion of equation \eqref{dy_RNC}. The first term of interest is the following

\begin{equation}
\int_{I[x,q]}\drm^dy\,\tau_{xy,f}^{dn}.
\end{equation}

As discussed in the main corpus of this work, on the integration domain $I[x,q]$ we define a set of orthonormal RNC $\{y^\mu\}$ centred at $O'$, in the middle of the $I[x,q]$ interval, and oriented along the $(q-x)^\mu$ vector, such that $y^\mu=(t_y,r_yn^i)$, with $t_y\in[-\frac{\tau'}{2},\frac{\tau'}{2}]$. The $n^i$ are radial space-like vectors normalised such that $n_in^i=1$. The extreme points are then $x^\mu=(-\tau'/2,0,...,0)$ and $q^\mu=(\tau'/2,0,...0)$. Lorentz invariance makes possible this choice of coordinates, such that the integration domain is spherically symmetric. To tackle the integrations, it is however useful to switch to null coordinates defined as 

 \begin{align}
v_y&=\frac{1}{\sqrt{2}}\left(\frac{\tau'}{2}+t_y+r_y\right),&  u_y&=\frac{1}{\sqrt{2}}\left(\frac{\tau'}{2}+t_y-r_y\right).
\end{align}

The integration measure becomes then 

\begin{equation}
 \int_{I[x,q]}\drm^dy=\int_0^{\tau'/\sqrt{2}}\drm v_y\,\int_0^{v_y}\drm u_y\,\left(\frac{v_y-u_y}{\sqrt{2}}\right)^{d-2}\int_{\mathcal{S}_{d-2}}\drm\Omega_{d-2}.
\end{equation}

The proper time $\tau_{xy,f}$ is simply given, in these coordinates, by $\tau_{xy,f}=\sqrt{2v_yu_y}$. Therefore,

\begin{align}
&\int_{I[x,q]}\drm^dy\,\tau_{xy,f}^{dn},\nonumber\\
&=\int_0^{\tau'/\sqrt{2}}\drm v_y\,\int_0^{v_y}\drm u_y\,\left(\frac{v_y-u_y}{\sqrt{2}}\right)^{d-2}\int_{\mathcal{S}_{d-2}}\drm\Omega_{d-2}(2v_y u_y)^{\frac{dn}{2}},\nonumber\\
&=\Omega_{d-2}\int_0^{\tau'/\sqrt{2}}\drm v_y\,\int_0^{v_y}\drm u_y\,\left(\frac{v_y-u_y}{\sqrt{2}}\right)^{d-2}(2v_y u_y)^{\frac{dn}{2}},\nonumber\\
&=\Omega_{d-2}\frac{n2^{-d}\Gamma(d-1)\Gamma\left(\frac{dn}{2}\right)}{(n+1)\Gamma\left(d+\frac{dn}{2}\right)}\tau'^{d(n+1)}.
\end{align}

Then, we have 

\begin{align}
\int_{\Delta I[x,q]}\drm^dy\,\tau_{xy,f}^{dn},
\end{align}

i.e. a term integrated over the corrections to the light cones bounding the integration domain. As outlined in the main text, the next terms can be treated Taylor expanding the integration. One can write

\begin{equation}
\int_a^{a+\varepsilon}\mathrm{d}r\,f(r)=f(a)\varepsilon+\mathcal{O}(\varepsilon^2),
\end{equation}

where $a=r^\pm(t,n)$ and 

\begin{equation}
\varepsilon^\pm=\left(\frac{\tau}{2}\mp t\right)\frac{R_{0i0j}n^in^j}{6}\left(\frac{\tau}{2}\right)^2.
\end{equation}

In this term, $\tau_{xy}=\tau_{xy,f}$ at the zeroth order in curvature corrections.

{\small
\begin{align}
\int_{\Delta I[x,q]}\drm^dy\,\tau_{xy,f}^{dn}=&\int\drm\Omega_{d-2}\Bigg[\int_{-\frac{\tau'}{2}}^0\drm t_y\int_{\frac{\tau'}{2}+t_y}^{\frac{\tau'}{2}+t_y+\varepsilon^-}\drm r_yr_y^{d-2}\tau_{xy,f}^{dn}+\int^{\frac{\tau'}{2}}_0\drm t_y\int_{\frac{\tau'}{2}-t_y}^{\frac{\tau'}{2}-t_y+\varepsilon^+}\drm r_yr_y^{d-2}\tau_{xy,f}^{dn}\Bigg],\nonumber\\
=&\int\drm\Omega_{d-2}\int^{\frac{\tau'}{2}}_0\drm t_y\Bigg[r_y^{d-2}\tau_{xy,f}^{dn}\left(\frac{\tau'}{2}- t_y\right)\frac{R_{0i0j}n^in^j}{6}\left(\frac{\tau'}{2}\right)^2\Bigg]_{r_y=\frac{\tau'}{2}-t_y},\nonumber\\
=&\Omega_{d-2}\int^{\frac{\tau'}{2}}_0\drm t_y \left(\frac{\tau'}{2}-t_y\right)^{d-1}\left(\left(\frac{\tau'}{2}+t_y\right)^2-\left(\frac{\tau'}{2}-t_y\right)^2\right)^{\frac{dn}{2}}\left(\frac{\tau'}{2}\right)^2\frac{R_{0'0'}}{6(d-1)},\nonumber\\
=&\Omega_{d-2}\int^{\frac{\tau'}{2}}_0\drm t_y \left(\frac{\tau'}{2}-t_y\right)^{d-1}\left(2t_y\tau'\right)^{\frac{dn}{2}}\left(\frac{\tau'}{2}\right)^2\frac{R_{0'0'}}{6(d-1)},\nonumber\\
=&\Omega_{d-2}\frac{2^{-3-d}\Gamma(d)\Gamma\left(1+\frac{dn}{2}\right)}{3(d-1)\Gamma\left(1+d+\frac{dn}{2}\right)}\tau'^{d(n+1)+2}R_{0'0'}.
\end{align}
}%

From the second to the third equality, we used the following identities 

\begin{align}
\int\mathrm{d}\Omega_dn^{i_1}n^{i_2}\dots n^{i_k}&=\frac{\Omega_d(d-1)!!}{(d+k-1)!!}\delta^{(k)}_{i_1i_2\dots i_k}, & &\text{if $k$ even}\nonumber\\
\int\mathrm{d}\Omega_dn^{i_1}n^{i_2}\dots n^{i_k}&=0, & &\text{if $k$ odd}.
\label{Omega}
\end{align}

The next term in the expansion is 

\begin{align}
&-\int_{I[x,q],f}\drm^dy\,\tau_{xy,f}^{dn}\frac{1}{6}R_{\mu\nu}y^\mu y^\nu,\nonumber\\
&=-\int_{I[x,q],f}\drm^dy\,\tau_{xy,f}^{dn}\frac{1}{6}\Bigg(R_{0'0'}t_y^2+R_{ij}r_y^2n^in^j\Bigg),\nonumber\\
&=-\frac{1}{6}\Omega_{d-2}\int_{I[x,q],f}\drm^2y\,\tau_{xy,f}^{dn}\Bigg(R_{0'0'}t_y^2+\tensor{R}{_i^i}\frac{r_y^2}{d-1}\Bigg),\nonumber\\
&=-\frac{1}{6}\Omega_{d-2}\int_{I[x,q],f}\drm^2y\,\tau_{xy,f}^{dn}\Bigg(R_{0'0'}t_y^2+(R_{0'0'}+R)\frac{r_y^2}{d-1}\Bigg),\nonumber\\
&=-\frac{1}{6}\Omega_{d-2}\int_{I[x,q],f}\drm^2y\,\tau_{xy,f}^{dn}\Bigg(R_{0'0'}t_y^2+(R_{0'0'}+R)\frac{r_y^2}{d-1}\Bigg),\nonumber\\
&=-\frac{1}{6}\Omega_{d-2}\int_0^{\tau'/\sqrt{2}}\drm v_y\,\int_0^{v_y}\drm u_y\,\left(\frac{v_y-u_y}{\sqrt{2}}\right)^{d-2}(2v_yu_y)^{\frac{dn}{2}}\Bigg(R_{0'0'}\left(\frac{v_y+u_y}{\sqrt{2}}-\frac{\tau'}{2}\right)^2\nonumber\\
&\quad+\frac{R_{0'0'}+R}{d-1}\left(\frac{v_y-u_y}{\sqrt{2}}\right)^{2}\Bigg),\nonumber\\
&=-\Omega_{d-2}\Bigg(\frac{2^{-4-d}(4+n(2+dn))\Gamma(d-1)\Gamma\left(1+\frac{dn}{2}\right)}{3(n+1)\Gamma\left(2+d+\frac{dn}{2}\right)}R_{0'0'}\nonumber\\
&\hspace{1.6cm}+\frac{2^{-2-d}\Gamma(d+1)\Gamma\left(1+\frac{dn}{2}\right)}{3(d-1)(2+d+dn)\Gamma\left(2+d+\frac{dn}{2}\right)}R\Bigg)\tau'^{d(n+1)+2},
\end{align}

Where we used again \eqref{Omega} and $R=-R_{0'0'}+\tensor{R}{_i^i}$. In the second line, terms in $R_{i0}$ were dropped because  they would have vanished in the angular integration.\\

Going on, one has the term coming from the expansion of the proper time

\begin{align}
\int_{I[x,q],f}\drm^dy\,\tau_{xy,f}^{dn-2}\frac{dn}{6}\left(\frac{\tau'}{2}\right)^2r_y^2R_{i0j0}n^in^j.
\end{align}

This comes from writing 

\begin{align}
\tau^2_{xy}=&-\eta_{\mu\nu}(y-x)^\mu(y-x)^\nu+\frac{1}{3}R_{\mu\alpha\nu\beta}y^\alpha y^\beta(y-x)^\mu(y-x)^\nu\nonumber\\
=&\tau_{xy,f}+\frac{1}{3}R_{\mu\alpha\nu\beta}y^\alpha y^\beta(y-x)^\mu(y-x)^\nu.
\end{align}

Contracting the Riemann tensor, we get

\begin{align}
&R_{\mu\alpha\nu\beta}y^\alpha y^\beta(y-x)^\mu(y-x)^\nu\nonumber,\\
=&R_{0000}t_y^2\left(\frac{\tau'}{2}+t_y\right)^2+R_{00ij}t_y\left(\frac{\tau'}{2}+t_y\right)r_y^2n^in^j+R_{0i0j}\left(\frac{\tau'}{2}+t_y\right)^2r_y^2n^in^j\nonumber \\
&-R_{0ij0}t_y\left(\frac{\tau'}{2}+t_y\right)r_y^2n^in^j+R_{i00j}t_y\left(\frac{\tau'}{2}+t_y\right)r_y^2n^in^j+R_{i0j0}t_y^2r_y^2n^in^j\nonumber\\
&+R_{ij00}t_y\left(\frac{\tau'}{2}+t_y\right)r_y^2n^in^j+R_{ijkl}r_y^4n^in^jn^kn^l,
  \label{TermRiemann}
\end{align}

where we already ignored the terms containing an odd number of $n^i$ vectors, which will vanish after angular integration, as from the properties \ref{Omega}. Equation \eqref{TermRiemann} can be further simplified noticing that $R_{0000}=0$ and

{\small
\begin{align}
    R_{\mu\alpha\nu\beta}y^\alpha y^\beta(y-x)^\mu(y-x)^\nu=&\left(R_{00ij}+R_{0ij0}+R_{i00j}+R_{ij00}\right)t_y\left(\frac{\tau'}{2}+t_y\right)r_y^2n^in^j\nonumber\\
    &+R_{0i0j}\left(\frac{\tau'}{2}+t_y\right)^2r_y^2n^in^j+R_{i0j0}t_y^2r_y^2n^in^j+R_{ijkl}r_y^4n^in^jn^kn^l\nonumber,\\
    =&-2R_{0i0j}t_y\left(\frac{\tau'}{2}+t_y\right)r_y^2n^in^j+R_{0i0j}\left(\frac{\tau'}{2}+t_y\right)^2r_y^2n^in^j\nonumber\\
    &+R_{i0j0}t_y^2r_y^2n^in^j+R_{ijkl}r_y^4n^in^jn^kn^l\nonumber,\\
    =&R_{0i0j}\left(-2t_y\left(\frac{\tau'}{2}+t_y\right)+\left(\frac{\tau'}{2}+t_y\right)^2+t_y^2\right)r_y^2n^in^j+R_{ijkl}r_y^4n^in^jn^kn^l\nonumber,\\
    =&R_{0i0j}\left(\frac{\tau'}{2}\right)^2r_y^2n^in^j+R_{ijkl}r_y^4n^in^jn^kn^l.
\end{align}
}%

From the first to the second equality, the first Bianchi identity has been applied, i.e.

\begin{equation}
    R_{\mu\nu\alpha\beta}+R_{\mu\alpha\beta\nu}+R_{\mu\beta\nu\alpha}=0.
    \label{Bianchi}
\end{equation}

Using \eqref{Omega} and \eqref{Bianchi} again, one can see that 

\begin{align}
    \int\drm \Omega_{d-2}R_{ijkl}r_y^4n^in^jn^kn^l=\Omega_{d-2}R_{ijkl}\frac{\delta^{(4)}_{ijkl}}{d^2-1}r_x^4=\Omega_{d-2}R_{ijkl}\frac{\delta_{ij}\delta_{kl}+\delta_{il}\delta_{jk}+\delta_{ik}\delta_{jl}}{d^2-1}r_y^4=0.
\end{align}

Therefore, 

\begin{align}
&\int_{I[x,q],f}\drm^dy\,\tau_{xy,f}^{dn-2}\frac{dn}{6}\left(\frac{\tau'}{2}\right)^2r_y^2R_{i0j0}n^in^j\nonumber\\
&=\frac{dn}{6(d-1)}\Omega_{d-2}\int_0^{\tau'/\sqrt{2}}\drm v_y\,\int_0^{v_y}\drm u_y\,\left(\frac{v_y-u_y}{\sqrt{2}}\right)^{d}(2v_yu_y)^{\frac{dn}{2}-1}\left(\frac{\tau'}{2}\right)^2\tensor{R}{_i_0^i_0},\nonumber\\
&=\Omega_{d-2}\frac{2^{-4-d}n\Gamma(d+1)\Gamma\left(\frac{dn}{2}\right)}{3(d-1)(n+1)\Gamma\left(1+d+\frac{dn}{2}\right)}\tau'^{d(n+1)+2}R_{0'0'}.
\end{align}

The second step was possible thanks to this equality

\begin{equation}
     R_{0i0j}\delta^{ij}=\tensor{R}{_0_i_0^i}=\tensor{R}{_0_i_0_i}=R_{00}.
\end{equation}

Finally, we are left with 

\begin{align}
&\int_{I[x,q],f}\drm^dy\,\tau_{xy,f}^{dn}\Bigg(-\frac{dn}{24(d+1)(d+2)}R\tau_{xy,f}^{2}+\frac{dn}{24(d+1)}R_{\mu\nu}(y-x)^\mu(y-x)^\nu\Bigg)\nonumber\\
=&\int_{I[x,q],f}\drm^dy\,\tau_{xy,f}^{dn}\frac{dn}{24(d+1)}\Bigg(-\frac{R}{(d+2)}\tau_{xy,f}^{2}+R_{0'0'}\left(t_y+\frac{\tau'}{2}\right)^2+R_{ij}r_y^2n^in^j\Bigg)\nonumber,\\
=&\Omega_{d-2}\int_0^{\tau'/\sqrt{2}}\drm v_y\,\int_0^{v_y}\drm u_y\,\left(\frac{v_y-u_y}{\sqrt{2}}\right)^{d-2}(2v_yu_y)^{\frac{dn}{2}}\frac{dn}{24(d+1)}\Bigg(-\frac{R}{(d+2)}(2v_yu_y)\nonumber\\
&+R_{0'0'}\left(\frac{v_y+u_y}{\sqrt{2}}\right)^2+\frac{R_{0'0'}+R}{d-1}\left(\frac{v_y-u_y}{\sqrt{2}}\right)^{2}\Bigg)\nonumber,\\
=&\Omega_{d-2}\Bigg(\frac{2^{-4-d}dn(2+d+dn)\Gamma(d)\Gamma\left(1+\frac{dn}{2}\right)}{3(d^2-1)\Gamma\left(2+d+\frac{dn}{2}\right)}R_{0'0'}\nonumber\\
&-\frac{2^{-4-d}(4+d(2+4n+d(-1+n(n+2))))\Gamma(d)\Gamma\left(1+\frac{dn}{2}\right)}{3(d^2-1)(d+2)(2+d+dn)\Gamma\left(2+d+\frac{dn}{2}\right)}R\Bigg)\tau'^{d(n+1)+2}.
\end{align}

The terms we have described here sum up to give equation \eqref{dy_final_RNC}. This must now be integrated over the causal domain from $p$ to $q$. The techniques used are analogous to the ones just described. We define a RNC system $\{x^\mu\}$ centred in $O$, i.e. in the middle of the $I[p,q]$ interval and oriented along the timelike geodesic from $p$ to $q$. Again, $x^\mu=(t_x, r_xn^i )$, with the radial vectors defined as before. The points $p$ and $q$ have now coordinates $(-\tau/2,0,...0)$ and $(\tau/2,0,...,0)$ respectively. We will now use

\begin{equation}
\tau'^2=\tilde{\tau}'^2+\frac{1}{3}R_{0i0j}r_x^2n^in^j\left(\frac{\tau}{2}\right)^2.
\end{equation}

Notice as well that $\tau$ is not affected by the RNC expansion as $p$ and $q$ have null radial components. \\

The main part of this integral is given by 

\begin{align}
&\Omega_{d-2}\frac{n2^{-d}\Gamma(d-1)\Gamma\left(\frac{dn}{2}\right)}{(n+1)\Gamma\left(d+\frac{dn}{2}\right)}\int_{I[p,q]}\drm^dx\,\sqrt{-g(x)}\tau'^{d(n+1)},\nonumber\\
=&\Omega_{d-2}\frac{n2^{-d}\Gamma(d-1)\Gamma\left(\frac{dn}{2}\right)}{(n+1)\Gamma\left(d+\frac{dn}{2}\right)}\int_{I[p,q]}\drm^dx\,\left(1-\frac{1}{6}R_{\mu\nu}x^\mu x^\nu\right)\left(\tilde{\tau}'^2+\frac{1}{3}R_{0i0j}r_x^2n^in^j\left(\frac{\tau}{2}\right)^2\right)^{\frac{d(n+1)}{2}},\nonumber\\
=&\Omega_{d-2}\frac{n2^{-d}\Gamma(d-1)\Gamma\left(\frac{dn}{2}\right)}{(n+1)\Gamma\left(d+\frac{dn}{2}\right)}\Bigg\{\int_{I[p,q],f}\drm^dx\,\tau_f'^{d(n+1)}+\int_{\Delta I[p,q]}\drm^dx\,\tau_f'^{d(n+1)}\nonumber\\
&+\int_{I[p,q],f}\drm^dx\,\tau_f'^{d(n+1)}\left(\frac{d(n+1)}{6}R_{0i0j}r_x^2n^in^j\left(\frac{\tau}{2}\right)^2\tau_f'^{-2}-\frac{1}{6}R_{\mu\nu}x^\mu x^\nu\right)\Bigg\}.
\end{align}

The second term in equation \eqref{dx_RNC} has to be integrated over $I[p,q],f$ after applying equation \eqref{rotation}, so to align $R_{0'0'}$ with the time-time direction of the external coordinate system. The third term in \eqref{dx_RNC} is directly integrated over $I[p,q],f$. Going term by term for completeness,

\begin{align}
&\Omega_{d-2}\frac{n2^{-d}\Gamma(d-1)\Gamma\left(\frac{dn}{2}\right)}{(n+1)\Gamma\left(d+\frac{dn}{2}\right)}\int_{I[p,q],f}\drm^dx\,\tau_f',\nonumber\\
=&\Omega_{d-2}^2\frac{n2^{-d}\Gamma(d-1)\Gamma\left(\frac{dn}{2}\right)}{(n+1)\Gamma\left(d+\frac{dn}{2}\right)}\int_0^{\tau/\sqrt{2}}\drm v_x\,\int_0^{v_x}\drm u_x\,\left(\frac{v_x-u_x}{\sqrt{2}}\right)^{d-2}(2v_xu_x)^{\frac{d(n+1)}{2}}\nonumber,\\
=&\Omega_{d-2}^2\frac{2^{-2d}\Gamma(d-1)^2\Gamma\left(\frac{dn}{2}(n+1)\right)\Gamma\left(1+\frac{dn}{2}\right)}{\Gamma\left(\frac{dn}{2}(n+3)\right)\Gamma\left(1+d+\frac{dn}{2}\right)}\tau^{d(n+2)}.
\end{align}

Next, 

\begin{align}
&\Omega_{d-2}\frac{n2^{-d}\Gamma(d-1)\Gamma\left(\frac{dn}{2}\right)}{(n+1)\Gamma\left(d+\frac{dn}{2}\right)}\int_{\Delta I[p,q]}\drm^dx\,\tau_f'\nonumber\\
=&\Omega_{d-2}\frac{n2^{-d}\Gamma(d-1)\Gamma\left(\frac{dn}{2}\right)}{(n+1)\Gamma\left(d+\frac{dn}{2}\right)}\int\drm\Omega_{d-2}\Bigg[\int_{-\frac{\tau}{2}}^0\drm t_x\int_{\frac{\tau}{2}+t_x}^{\frac{\tau}{2}+t_x+\varepsilon^-}\drm r_xr_x^{d-2}\tau_f'+\int^{\frac{\tau}{2}}_0\drm t_x\int_{\frac{\tau}{2}-t_x}^{\frac{\tau}{2}-t_x+\varepsilon^+}\drm r_xr_x^{d-2}\tau_f'\Bigg]\nonumber,\\
=&\Omega_{d-2}^2\frac{n2^{-d}\Gamma(d-1)\Gamma\left(\frac{dn}{2}\right)}{(n+1)\Gamma\left(d+\frac{dn}{2}\right)}\int_{-\frac{\tau'}{2}}^0\drm t_x \left(\frac{\tau}{2}+t_x\right)^{d-1}\left(\left(\frac{\tau}{2}-t_x\right)^2-\left(\frac{\tau}{2}+t_x\right)^2\right)^{\frac{d(n+1)}{2}}\left(\frac{\tau}{2}\right)^2\frac{R_{00}}{6(d-1)}\nonumber,\\
=&\Omega_{d-2}^2\frac{2^{-3-2d}\Gamma(d-1)^2\Gamma\left(\frac{dn}{2}(n+1)\right)\Gamma\left(1+\frac{dn}{2}\right)}{3\Gamma\left(d+\frac{dn}{2}\right)\Gamma\left(1+\frac{d}{2}(n+3)\right)}\tau^{d(n+2)+2}R_{00}.
\end{align}

The next term is given by

\begin{align}
&\Omega_{d-2}\frac{n2^{-d}\Gamma(d-1)\Gamma\left(\frac{dn}{2}\right)}{(n+1)\Gamma\left(d+\frac{dn}{2}\right)}\int_{I[p,q],f}\drm^dx\,\tau_f'^{d(n+1)-2}\frac{d(n+1)}{6}R_{0i0j}r_x^2n^in^j\left(\frac{\tau}{2}\right)^2\nonumber,\\
=&\Omega_{d-2}^2\frac{n2^{-d}\Gamma(d-1)\Gamma\left(\frac{dn}{2}\right)}{(n+1)\Gamma\left(d+\frac{dn}{2}\right)}\int_0^{\tau/\sqrt{2}}\drm v_x\,\int_0^{v_x}\drm u_x\,\left(\frac{v_x-u_x}{\sqrt{2}}\right)^{d}(2v_xu_x)^{\frac{d(n+1)}{2}-1}\frac{d(n+1)}{6(d-1)}R_{00}\left(\frac{\tau}{2}\right)^2\nonumber,\\
=&\Omega_{d-2}^2\frac{2^{-5-2d}n\Gamma(d-1)^2\Gamma\left(\frac{dn}{2}\right)\Gamma\left(\frac{d}{2}(n+1)\right)}{3(d-1)^2\Gamma\left(1+d+\frac{dn}{2}\right)\Gamma\left(1+\frac{d}{2}(n+3)\right)}\tau^{d(n+2)+2}R_{00}.
\end{align}

Finally,

\begin{align}
&-\Omega_{d-2}\frac{n2^{-d}\Gamma(d-1)\Gamma\left(\frac{dn}{2}\right)}{(n+1)\Gamma\left(d+\frac{dn}{2}\right)}\int_{I[p,q],f}\drm^dx\,\tau_f'\frac{1}{6}R_{\mu\nu}x^\mu x^\nu,\nonumber\\
=&-\Omega_{d-2}\frac{n2^{-d}\Gamma(d-1)\Gamma\left(\frac{dn}{2}\right)}{(n+1)\Gamma\left(d+\frac{dn}{2}\right)}\int_{I[p,q],f}\drm^dx\,\tau_f'\frac{1}{6}\left(R_{00}t_x^2+(R_{00}+R)\frac{r_x^2}{d-1}\right)\nonumber,\\
=&-\Omega_{d-2}\Bigg(R_{00}\frac{2^{-5-2d}d(d(n+1^2+2(n+3)))\Gamma(d-1)^2\Gamma\left(1+\frac{dn}{2}\right)\Gamma\left(\frac{d}{2}(n+1)\right)}{3\Gamma\left(1+d+\frac{dn}{2}\right)\Gamma\left(2+\frac{d}{2}(n+3)\right)}\nonumber\\
&+R\frac{2^{-3-2d}n\Gamma(d+1)^2\Gamma\left(\frac{dn}{2}\right)\Gamma\left(\frac{d}{2}(n+1)\right)}{3(d-1)^2(2+d(n+2))\Gamma\left(d+\frac{dn}{2}\right)\Gamma\left(2+\frac{d}{2}(n+3)\right)}\Bigg)\tau^{d(n+2)+2}.
\end{align}

We omit here the second and third term in equation \eqref{dx_RNC} for sake of simplicity, as they result in quite lengthy terms. However, their evaluation is analogous to the terms previously discussed.

\bibliographystyle{JHEP}
\bibliography{References}

\end{document}